\documentclass[epjST]{svjour}
\usepackage{graphics}
\usepackage{latexsym}
\usepackage{amssymb}
\usepackage{amsmath}
\usepackage{dsfont}
\usepackage{epsfig}
\newcommand{\ud}{\mathrm{d}}

\newcommand{\uTr}{\mathrm{Tr}}

\newcommand{\uk}{\mathbf{k}}

\begin{document}

\title{The partonic structure of the nucleon from generalized transverse momentum-dependent parton distributions}
\author{B. Pasquini\inst{1}\fnmsep\thanks{\email{pasquini@pv.infn.it} }\and C. Lorc\'e \inst{2} \fnmsep\thanks{\email{lorce@ipno.in2p3.fr} } }
\institute{Dipartimento di Fisica, Universit\`a degli Studi di Pavia, Pavia, Italy\\
and Istituto Nazionale di Fisica Nucleare, Sezione di Pavia, Pavia, Italy \and IPNO, Universit\'e Paris-Sud, CNRS/IN2P3, 91406 Orsay, France\\
and LPT, Universit\'e Paris-Sud, CNRS, 91406 Orsay, France}
\abstract{
We discuss the general formalism for the calculation in light-front quark models of the fully unintegrated, off-diagonal quark-quark correlator of the nucleon, parametrized in terms of generalized transverse momentum dependent parton distributions (GTMDs). By taking specific limits or projections, these GTMDs yield various transverse-momentum dependent and generalized parton distributions, thus providing a unified framework to simultaneously model different observables. The corresponding distributions in impact-parameter space are the Wigner functions which provide multidimensional images of the
quark distributions in phase space.
 We present results within a light-front constituent quark model, discussing some of the complementary aspects encoded in the 
different distributions and the relation to the quark orbital angular momentum of the proton. } 
\maketitle
\section{Introduction}
\label{intro}
\begin{figure}[t!]
\begin{center}
\epsfig{file=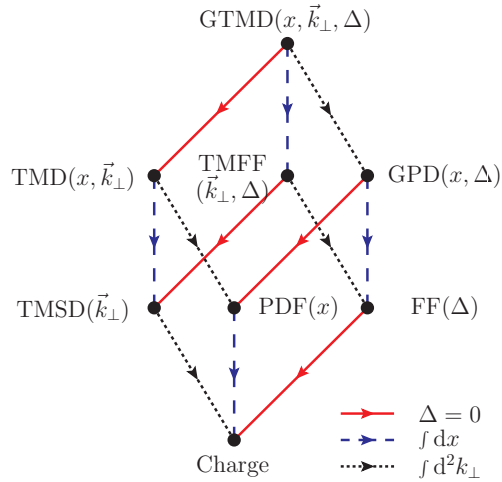,  width=0.5\columnwidth}
\end{center}
\caption{\footnotesize{Representation of the projections of the GTMDs into 
parton distributions and form factors.
The arrows correspond to different reductions in the hadron and quark momentum
space: the solid (red) arrows  give the forward limit in the hadron momentum,
the dotted (black) arrows correspond to integrating over the quark 
transverse-momentum and the dashed (blue) arrows project out the longitudinal momentum of quarks.
The different objects resulting from these links are explained in the text.}}
\label{fig1}
\end{figure}

Parton distributions entering many hard and exclusive processes play a
key role to describe the nonperturbative structure of hadrons.
The most complete information is contained
in the generalized transverse momentum dependent parton distributions (GTMDs)~\cite{Meissner:2009ww,Meissner:2008ay,Lorce:2011dv} which parametrize the unintegrated off-diagonal quark-quark correlator, depending on the quark longitudinal  and transverse momentum,  $k^+$ and $\vec k_\perp$, respectively, and on the 4-momentum $\Delta$ which is transferred by the probe to the hadron; for a classification see Refs.~\cite{Meissner:2009ww,Meissner:2008ay}.
The GTMDs give the full one-quark density matrix in the momentum space and 
  reduce to different
parton distributions and form factors as is shown in Fig.~\ref{fig1}. 

The different arrows in this
figure represent particular projections in the hadron and quark momentum space, and give
the links between the matrix elements of different reduced density matrices.
Such matrix elements can in turn be parametrized in terms of generalized parton distributions
(GPDs), transverse-momentum dependent parton distributions (TMDs) and
form factors (FFs). These are the quantities which enter the description of various
exclusive (GPDs), semi-inclusive (TMDs), and inclusive (PDFs) deep inelastic scattering
processes, or parameterize elastic scattering processes (FFs). At leading twist, there are
sixteen complex GTMDs, which are defined in terms of the independent polarization states
of quarks and hadron. In the forward limit $\Delta= 0$ they reduce to eight TMDs which depend
on the longitudinal momentum fraction $x$ and transverse momentum ~$\vec k _\perp$ of quarks,
and therefore give access to the three-dimensional picture of the hadrons in momentum
space. On the other hand, the integration over ~$\vec k_\perp$ of the GTMDs leads to eight GPDs
which are probability amplitudes related to the off-diagonal matrix elements of the parton
density matrix in the longitudinal momentum space. The common limit of TMDs and GPDs is
given by the standard parton distribution functions (PDFs), related to the diagonal matrix
elements of the longitudinal-momentum density matrix for different polarization states of
quarks and hadron. The integration over $x$ of the GTMDs leads to a bilocal operator restricted to the
plane transverse to the light-front direction and brings to the lower plane of the box in Fig.~\ref{fig1}.
The off-forward matrix elements of this operator can be parametrized in terms of so-called
transverse-momentum dependent form factors (TMFFs). Starting from the TMFFs, we
can follow the same path as in the case of the GTMDs, and at each vertex of the basis of
the box of Fig.~\ref{fig1} we find the restricted version of the operator defining the distributions
in the upper plane. Therefore, integrating out the dependence on the quark transverse
momentum, we encounter matrix elements parametrized in terms of form factors (FFs),
while the forward limit of TMFFs leads to transverse-momentum dependent spin densities
(TMSD). Both FFs and TMSDs have the charges as common limit.
\\
After
appropriate Fourier transform, the GTMDs can be interpreted as Wigner or phase-space distributions
~\cite{Ji:2003ak,Belitsky:2003nz,Belitsky:2005qn,Lorce:2011kd,Lorce:2011ni},
giving access to the correlations between quark momentum and transverse position.
The Wigner distributions reduce to the Fourier transform of the GPDs in impact-parameter space (or impact-parameter dependent distributions)  after integration over the quark transverse momentum, and, upon further integration over the longitudinal quark momentum, 
to the charge densities in the transverse coordinate plane.
\\
Although a variety of models has been employed to explore separately the different
observables related to GTMDs, a unifying formalism for modeling the GTMDs has been presented only recently~\cite{Lorce:2011dv}.
In the following, we will review some of the results discussed in Ref.~\cite{Lorce:2011dv}, using the language of light-front wave functions
(LFWFs) and  focusing on the three-quark (3Q) contribution.
In Sect.~\ref{sec:1} we present the formal derivation of
the LFWF overlap representation of the quark contribution to GTMDs, specializing the
results to two light-front quark models, namely the light-front chiral quark-soliton model (LF$\chi$QSM) and
the light-front constituent quark model (LFCQM).  
In Sect.~\ref{sec:2} we introduce the Wigner distributions,  discussing the case of unpolarized quarks in the longitudinally polarized nucleon 
and its relation to the quark OAM.
Then, in Sects.~\ref{sec:3} and \ref{sec:4},  we discuss 
 some of the complementary aspects encoded in the GPDs and the TMDs, in particular with regards to the  information on the quark OAM.  Concluding remarks are given in the final section.
  
 \section{Quark-quark Correlator}
\label{sec:1}
\begin{figure}[ht]
\begin{center}
\epsfig{file=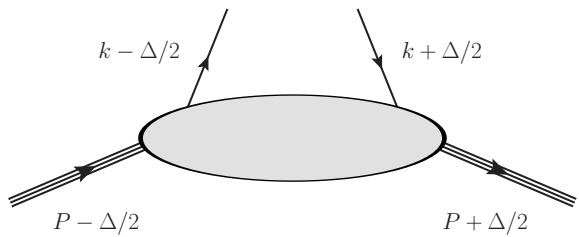,  width=0.6\columnwidth}
\end{center}
\caption{Kinematics for the fully-unintegrated quark-quark correlator.}
\label{fig2}
\end{figure}

The fully-unintegrated quark-quark correlator $\tilde W$ for a spin-$1/2$ hadron is defined as~\cite{Meissner:2009ww,Meissner:2008ay} \begin{equation}\label{gencorr}
\tilde W^{[\Gamma]}_{\Lambda'\Lambda}(P,k,\Delta,N;\eta)=\frac{1}{2}\int\frac{\ud^4z}{(2\pi)^4}\,e^{ik\cdot z}\,\langle p',\Lambda'|\overline\psi(-\tfrac{1}{2}z)\Gamma\,\mathcal W\,\psi(\tfrac{1}{2}z)|p,\Lambda\rangle.
\end{equation}
This correlator is a function of the initial and final hadron light-front helicities $\Lambda$ and $\Lambda'$, the average hadron and quark four-momenta $P=(p'+p)/2$ and $k$, and the four-momentum transfer to the hadron $\Delta=p'-p$ (see Fig.~\ref{fig2} for the kinematics).
 The superscript $\Gamma$ stands for any element of the basis $\{\mathds 1,\gamma_5,\gamma^\mu,\gamma^\mu\gamma_5,i\sigma^{\mu\nu}\}$ in Dirac space. A Wilson line $\mathcal W\equiv\mathcal W(-\tfrac{1}{2}z,\tfrac{1}{2}z|n)$ ensures the color gauge invariance of the correlator, connecting the points $-\tfrac{1}{2}z$ and $\tfrac{1}{2}z$ \emph{via} the intermediary points $-\tfrac{1}{2}z+\infty\cdot n$ and $\tfrac{1}{2}z+\infty\cdot n$ by straight lines. This induces a dependence of the Wilson line on the light-front direction $n$. Since any rescaled four-vector $\alpha n$ with some positive parameter $\alpha$ could be used to specify the Wilson line, the correlator actually only depends on the four-vector $N=\frac{M^2n}{P\cdot n}$, where $M$ is the hadron mass. The parameter $\eta=\textrm{sign}(n^0)$ gives the sign of the zeroth component of $n$, \emph{i.e.} indicates whether the Wilson line is future-pointing ($\eta=+1$) or past-pointing ($\eta=-1$). 

Since the parton light-front energy $k^-$ is particularly difficult to access in high-energy experiments, the relevant correlators are actually obtained from the  $k^-$ integrated version of Eq.~\eqref{gencorr}, setting all the fields at the same light-front time $z^+=0$:
\begin{equation}\label{GTMDcorr}
\begin{split}
W^{[\Gamma]}_{\Lambda'\Lambda}(P,x,\vec k_\perp,\Delta,N;\eta)&=\int\ud k^-\,\tilde W^{[\Gamma]}_{\Lambda'\Lambda}(P,k,\Delta,N;\eta)\\
&=\frac{1}{2}\int\frac{\ud z^-\,\ud^2z_\perp}{(2\pi)^3}\,e^{ik\cdot z}\,\langle p',\Lambda'|\overline\psi(-\tfrac{1}{2}z)\Gamma\,\mathcal W\,\psi(\tfrac{1}{2}z)|p,\Lambda\rangle\Big|_{z^+=0},
\end{split}
\end{equation}
where we used for a generic four-vector $a^\mu=[a^+,a^-,\vec a_\perp]$ the light-front components $a^\pm=(a^0\pm a^3)/\sqrt{2}$ and the transverse components $\vec a_\perp=(a^1,a^2)$, and $x=k^+/P^+$ is the average fraction of longitudinal momentum carried by the quark. A complete parametrization of this object in terms of GTMDs has been achieved in~\cite{Meissner:2009ww,Meissner:2008ay}. 

\subsection{Overlap Representation}
Following
 the lines of~\cite{Diehl:2000xz,Brodsky:2000xy}, we obtain in the light-front gauge $A^+=0$ an overlap representation for the correlator~\eqref{GTMDcorr} at the twist-two level restricted to the 3Q Fock sector\footnote{Quark flavor and color indices have been omitted for clarity. In the processes considered here the flavor and color of a given quark remain unchanged.} 
 \begin{eqnarray}
&&W^{[\Gamma]}_{\Lambda'\Lambda}(P,x,\vec k_\perp,\Delta,N;\eta)\nonumber\\
&&=\frac{1}{\sqrt{1-\xi^2}}\sum_{\lambda'_i,\lambda_i}\int[\ud x]_3\,[\ud^2k_\perp]_3\,\Delta(\tilde k)\,\psi^*_{\Lambda'\beta'}(r')\,\psi_{\Lambda\beta}(r)\prod_{i=1}^3 M^{\lambda'_i\lambda_i},\label{overlap}
\end{eqnarray}
where the integration measures are defined as
\begin{equation}
[\ud x]_3\equiv\left[\prod_{i=1}^3\ud x_i\right]\delta\!\!\left(1-\sum_{i=1}^3x_i\right),\qquad[\ud^2k_\perp]_3\equiv\left[\prod_{i=1}^3\frac{\ud^2k_{i\perp}}{2(2\pi)^3}\right]2(2\pi)^3\,\delta^{(2)}\!\!\left(\sum_{i=1}^3\vec k_{i\perp}\right).
\end{equation}
Furthermore, in Eq.~\eqref{overlap} the function $\Delta(\tilde k)=3\,\Theta(x_1)\,\delta(x-x_1)\,\delta^{(2)}(\vec k_\perp-\vec k_{1\perp})$ selects the active quark average momentum (we choose to label the active quark with $i=1$ and the spectator quarks with $j=2,3$). The 3Q LFWF $\psi_{\Lambda\beta}(r)$ depends on the momentum coordinates $\tilde k_i=(y_i,\kappa_{i\perp})$ of the quarks relative to the hadron momentum (collectively indicated by $r$), and the index $\beta$ which  stands for the set of the quark light-front helicities $\{\lambda_i\}$. The transition from the initial quark light-front helicity $\lambda_i$ to the final one $\lambda'_i$ is described by a complex-valued $2\times 2$ matrix $M^{\lambda'_i\lambda_i}$. In particular, we have for the spectator quarks $M^{\lambda'_j\lambda_j}=\delta^{\lambda'_j\lambda_j}$. For the active quark, the matrix $M^{\lambda'_1\lambda_1}$ depends on the twist-two Dirac structure $\Gamma_\text{twist-2}=\{\gamma^+,\gamma^+\gamma_5,i\sigma^{1+}\gamma_5,i\sigma^{2+}\gamma_5\}$ used in the correlator, see \emph{e.g.}~\cite{Boffi:2002yy,Boffi:2003yj,Pasquini:2005dk}.
We choose to work in an infinite momentum frame such that $P^+$ is large, $\vec P_\perp=\vec 0_\perp$ and $\Delta\cdot P=0$. The four-momenta involved are then
\begin{equation}
\begin{aligned}
P&=\left[P^+,\frac{M^2+\tfrac{\Delta_\perp^2}{4}}{2(1-\xi^2)P^+},\vec 0_\perp\right],\qquad
&\Delta&=\left[-2\xi P^+,\xi\,
\frac{M^2+\tfrac{\Delta_\perp^2}{4}}{(1-\xi^2)P^+},\vec\Delta_\perp\right],\\
k&=\left[xP^+,k^-,\vec k_\perp\right],&n&=\left[0,\pm 1,\vec 0_\perp\right].
\end{aligned}
\end{equation}
Note that the form used for $n$ is not the most general one, but leads to an appropriate definition of TMDs for semi-inclusive deep inelastic and Drell-Yan processes. For the active and spectator quarks the initial and final momentum coordinates are then
\begin{equation}
\begin{aligned}
\tilde k_1&=\left(\frac{x+\xi}{1+\xi},\vec k_\perp-\frac{1-x}{1+\xi}\,\frac{\vec\Delta_\perp}{2}\right),\qquad&\tilde k'_1&=\left(\frac{x-\xi}{1-\xi},\vec k_\perp+\frac{1-x}{1-\xi}\,\frac{\vec\Delta_\perp}{2}\right),\\
\tilde k_j&=\left(\frac{x_j}{1+\xi},\vec k_{j\perp}+\frac{x_j}{1+\xi}\,\frac{\vec\Delta_\perp}{2}\right),&\tilde k'_j&=\left(\frac{x_j}{1-\xi},\vec k_{j\perp}-\frac{x_j}{1-\xi}\,\frac{\vec\Delta_\perp}{2}\right).
\end{aligned}
\end{equation}
So far, the exact 3Q LFWF derived directly from the QCD Lagrangian is not known. Nevertheless, we can try to reproduce the gross features of hadron structure 
at low  scales using constituent quark models. Many models exist on the market based on the concept of constituent quarks. However only a few incorporate consistently relativistic effects. We focus here on two such models, the light-front constituent quark model (LFCQM)~\cite{Boffi:2002yy,Boffi:2003yj,Pasquini:2005dk} and the light-front chiral quark-soliton model (LF$\chi$QSM)~\cite{Petrov:2002jr,Diakonov:2005ib,Lorce:2007as,Lorce:2007fa}. 
However, the formalism can be easily generalized to other quark models as explained in Refs.~\cite{Lorce:2011zta,Pasquini:2012jm}.
\\
The LFWFs used in LFCQM and in $LF\chi$QSM have a very similar structure, given by
\begin{equation}\label{LCWF}
\psi_{\Lambda\beta}(r)=\Psi(r)\sum_{\sigma_i}\Phi_\Lambda^{\sigma_1\sigma_2\sigma_3}\prod_{i=1}^3 D_{\lambda_i\sigma_i}(\tilde k_i),
\end{equation}
where $\Psi(r)$ is a global symmetric momentum wave function, $\Phi_\Lambda^{\sigma_1\sigma_2\sigma_3}$ is the $SU(6)$ spin-flavor wave function, and $D(\tilde k)$ is an $SU(2)$ matrix connecting light-front helicity $\lambda_i$ and canonical spin $\sigma_i$
\begin{equation}\label{generalizedMelosh}
D(\tilde k)=\frac{1}{|\vec K|}\begin{pmatrix}K_z&K_L\\-K_R&K_z\end{pmatrix},\qquad K_{R,L}=K^1\pm iK^2.
\end{equation}
The explicit expressions for the momentum wave function $\Psi(r)$ 
in Eq.~\eqref{LCWF} and the vector $\vec K$ in Eq.~\eqref{generalizedMelosh} 
in LFCQM read
\begin{eqnarray}
\Psi(r)=2(2\pi)^3\sqrt{\frac{\omega_1\omega_2\omega_3}{x_1x_2x_3\mathcal M_0}}\,\frac{\mathcal N}{(\mathcal M_0^2+\beta^2)^\gamma},\nonumber\\
K_z=m+y\mathcal M_0,\qquad \vec K_\perp=\boldsymbol{\kappa}_\perp,\qquad \kappa_z=y\mathcal M_0-\omega,
\end{eqnarray}
where $\mathcal N$ is a normalization factor, $\mathcal M_0=\sum_i\omega_i$ is the free invariant mass, $\omega_i$ is the free energy of quark $i$, $m$ is the constituent quark mass, and $\beta,\gamma$ are  model parameters fitted to reproduce the anomalous magnetic moments of the nucleon~\cite{Pasquini:2007iz}.
On the other hand, within the   $LF\chi$QSM one has
\begin{equation}
\Psi(r)=\mathcal N\prod_{i=1}^3|\vec K_i|,\qquad K_z=h+\frac{\kappa_z}{|\boldsymbol{\kappa}|}\,j,\qquad \vec K_\perp=\frac{\boldsymbol{\kappa}_\perp}{|\boldsymbol{\kappa}|}\,j,\qquad \kappa_z=y\mathcal M_N-E_\text{lev},
\end{equation}
where $\mathcal M_N$ is the soliton mass, $E_\text{lev}$ is the energy of the discrete level in the spectrum, and $h,j$ are the upper and lower components of the Dirac spinor describing this discrete level.

For further convenience we introduce the tensor correlator
\begin{equation}
\label{master}
W^{\mu\nu}\equiv\frac{1}{2}\uTr\left[\bar\sigma^\mu W^\nu\right]=\frac{1}{2}\sum_{\Lambda'\Lambda}(\bar\sigma^\mu)^{\Lambda\Lambda'}W^\nu_{\Lambda'\Lambda},
\end{equation}
where $W^\nu_{\Lambda'\Lambda}\equiv\left(W^{[\gamma^+]}_{\Lambda'\Lambda},W^{[i\sigma^{1+}\gamma_5]}_{\Lambda'\Lambda},W^{[i\sigma^{2+}\gamma_5]}_{\Lambda'\Lambda},W^{[\gamma^+\gamma_5]}_{\Lambda'\Lambda}\right)$ and $\bar\sigma^\mu=(\mathds{1},\boldsymbol{ \sigma})$ with $\sigma_i$ the Pauli matrices. We now use the LFWF given by Eq.~\eqref{LCWF} and write the overlap representation of the correlator tensor $W^{\mu\nu}$ as
\begin{equation}\label{master1}
W^{\mu\nu}(P,x,\vec k_\perp,\Delta,N;\eta)=\frac{1}{\sqrt{1-\xi^2}}\int[\ud x]_3\,[\ud^2k_\perp]_3\,\Delta(\tilde k)\,\Psi^*(r')\,\Psi(r)\,\mathcal A^{\mu\nu}(r',r),
\end{equation}
where $\mathcal A^{\mu\nu}(r',r)$ stands for 
\begin{equation}\label{master2}
\mathcal A^{\mu\nu}(r',r)=A\,O_1^{\mu\nu}\left(l_2\cdot l_3\right)+B\left[l_2^\mu\left(l_3\cdot O_1\right)^\nu+l_3^\mu\left(l_2\cdot O_1\right)^\nu\right].
\end{equation}
In Eq.~\eqref{master2}, $l^\mu_j=O^{\mu0}_j$ and the matrix $O^{\mu\nu}$ is given by
\begin{equation}\label{spinhelicity}
O^{\mu\nu}=\frac{1}{|\vec K'||\vec K|}\begin{pmatrix}
\vec K'\cdot\vec K&i\left(\vec K'\times\vec K\right)_x&i\left(\vec K'\times\vec K\right)_y&-i\left(\vec K'\times\vec K\right)_z\\
i\left(\vec K'\times\vec K\right)_x&\vec K'\cdot\vec K-2K'_xK_x&-K'_xK_y-K'_yK_x&K'_xK_z+K'_zK_x\\
i\left(\vec K'\times\vec K\right)_y&-K'_yK_x-K'_xK_y&\vec K'\cdot\vec K-2K'_yK_y&K'_yK_z+K'_zK_y\\
i\left(\vec K'\times\vec K\right)_z&-K'_zK_x-K'_xK_z&-K'_zK_y-K'_yK_z&-\vec K'\cdot\vec K+2K'_zK_z
\end{pmatrix}.
\end{equation}
The tensor correlator $W^{\mu\nu}$ in Eq.~(\ref{master1}) has two indices. The index 
$\mu$ refers to the transition in terms of hadron light-front helicity, while the index $\nu$ refers to the transition in terms of the active quark light-front helicity. For example, the components $W^{00}$ and $W^{03}$ correspond to the matrix elements of the $\gamma^+$ and $\gamma^+\gamma_5$ operators in the case of an unpolarized hadron, respectively.
Equation~\eqref{master1} gives the explicit expression for  the tensor correlator in terms of the overlap of initial $\Psi(r)$ and final $\Psi^*(r')$ symmetric (instant-form) momentum wave functions with the tensor $\mathcal A^{\mu\nu}(r',r)$ for a fixed average momentum of the active quark $\Delta(\tilde k)$. 
The tensor $\mathcal A^{\mu\nu}(r',r)$  contains the spin-flavor structure 
derived from the overlap of the three initial and final quarks.
Taking into account the possible couplings of the helicities of the active and spectator quarks to give the hadron helicity, the coefficient 
$A$ and $B$ in Eq.~(\ref{master}) for SU(6)  spin-flavor wave functions
are \begin{equation}
A^p_u=4,\qquad B^p_u=1,\qquad A^p_d=-1,\qquad B^p_d=2.
\end{equation}
Furthermore, the matrix $O^{\mu\nu}$ in Eq.~(\ref{master1}) describes the overlap 
of the initial and final quark state. The columns are labeled by the index $\nu$ which indicates the type of transition in terms of quark light-front helicity. The rows are labeled by the index $\mu$ which indicates the type of transition in terms of quark canonical spin.
This matrix reduces to
$l^\mu_i=O^{\mu0}_i$ for the spectator quarks, 
since in this case the light-front helicity is conserved.
\section{Wigner distributions}
\label{sec:2}
By performing a Fourier transform of the GTMDs to the impact-parameter space, we  obtain quark distributions which are naturally interpreted as Wigner distributions~\cite{Ji:2003ak,Belitsky:2003nz,Lorce:2011kd}
\begin{equation}\label{wigner}
\rho^{[\Gamma]}_{\Lambda'\Lambda}(x,\vec k_\perp,\vec b_\perp,n)\equiv\int\frac{\ud^2\Delta_\perp}{(2\pi)^2}\,e^{-i\vec\Delta_\perp\cdot\vec b_\perp}\,W^{[\Gamma]}_{\Lambda'\Lambda}(P,x,\vec k_\perp,\Delta,n).
\end{equation}
Although the GTMDs are in general complex-valued functions, their two-dimensional Fourier transforms are always real-valued functions, in accordance with their interpretation as phase-space distributions.
We note that, like in the usual quantum-mechanical Wigner distributions, $\vec b_\perp$
and $\vec k_\perp$ are not Fourier conjugate variables. However, they are subjected to Heisenberg's
uncertainty principle because the corresponding quantum-mechanical operators
do not commute $[\hat{\vec b}_\perp,\hat{\vec k}_\perp] \ne 0$. As a consequence, the Wigner functions can not have a
strict probabilistic interpretation. There are in total 16 Wigner functions at twist-two
level, corresponding to all the 16 possible configurations of nucleon and quark polarizations.
Here we will discuss only one particular case, namely the distortion in the
distribution of unpolarized quarks due to the longitudinal polarization of the nucleon
$\rho^q_{LU} = \rho^{[\gamma^+]q}(\vec b_\perp,\vec k_\perp, x,+\vec e_z)-\rho^{[\gamma^+]q}(\vec b_\perp,\vec k_\perp, x,-\vec e_z)$ which has a close connection with the
quark orbital angular momentum (OAM). Other configurations for the quark and nucleon
polarizations can be found in Ref.~\cite{Lorce:2011kd} .

\begin{figure}[th!]
	\centering
		\includegraphics[width=.48\textwidth]{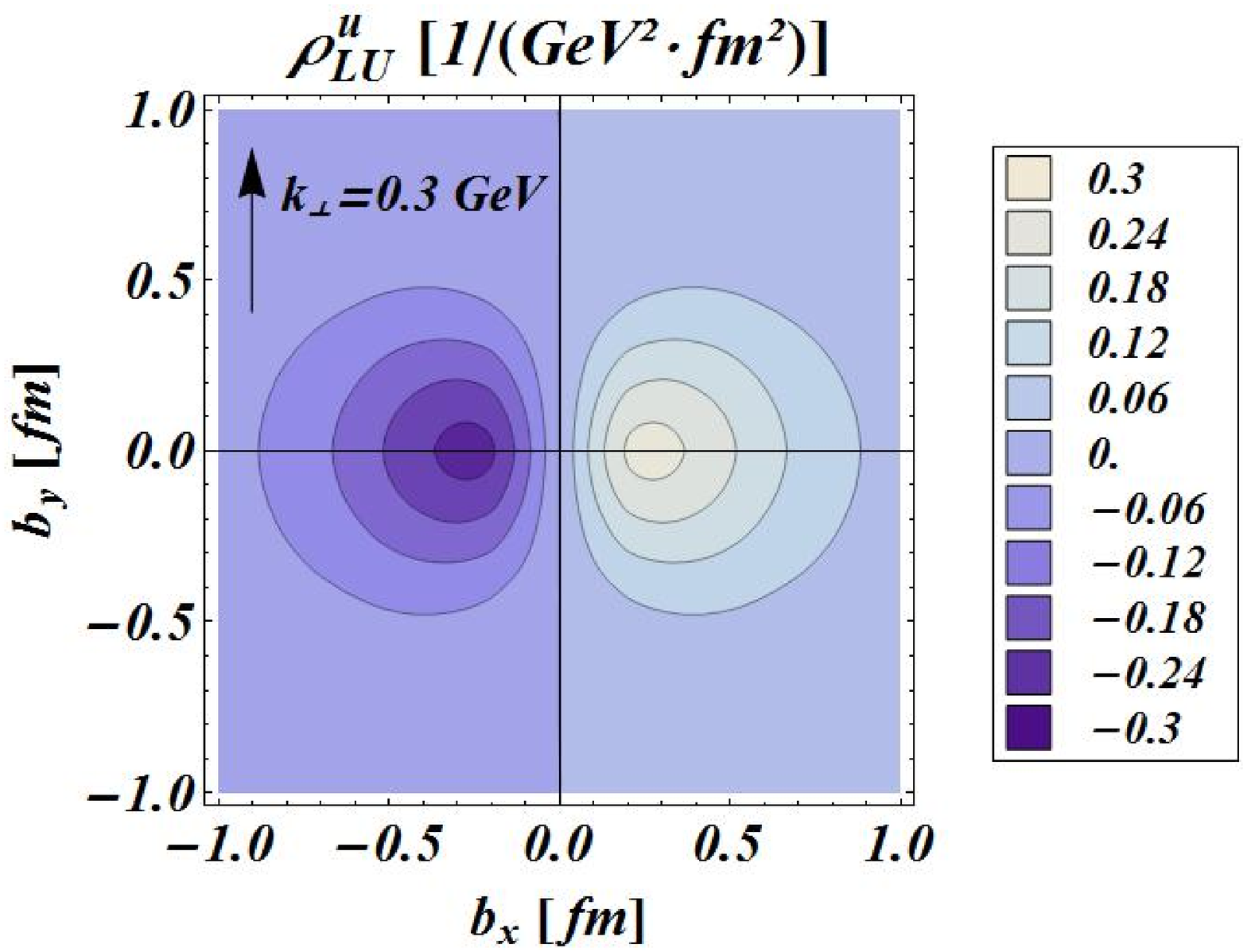}
		\includegraphics[width=.48\textwidth]{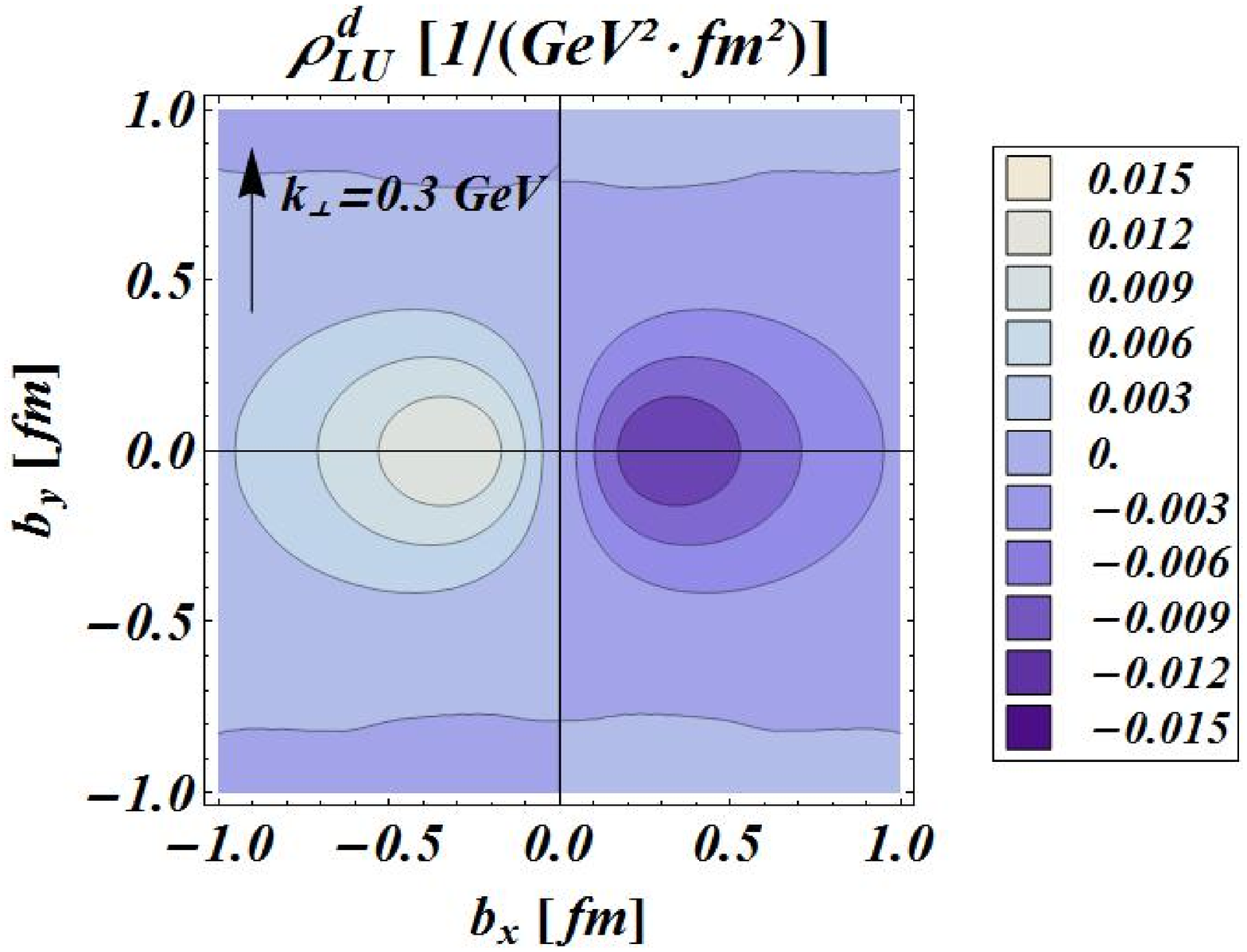}
                     \includegraphics[width=.42\textwidth]{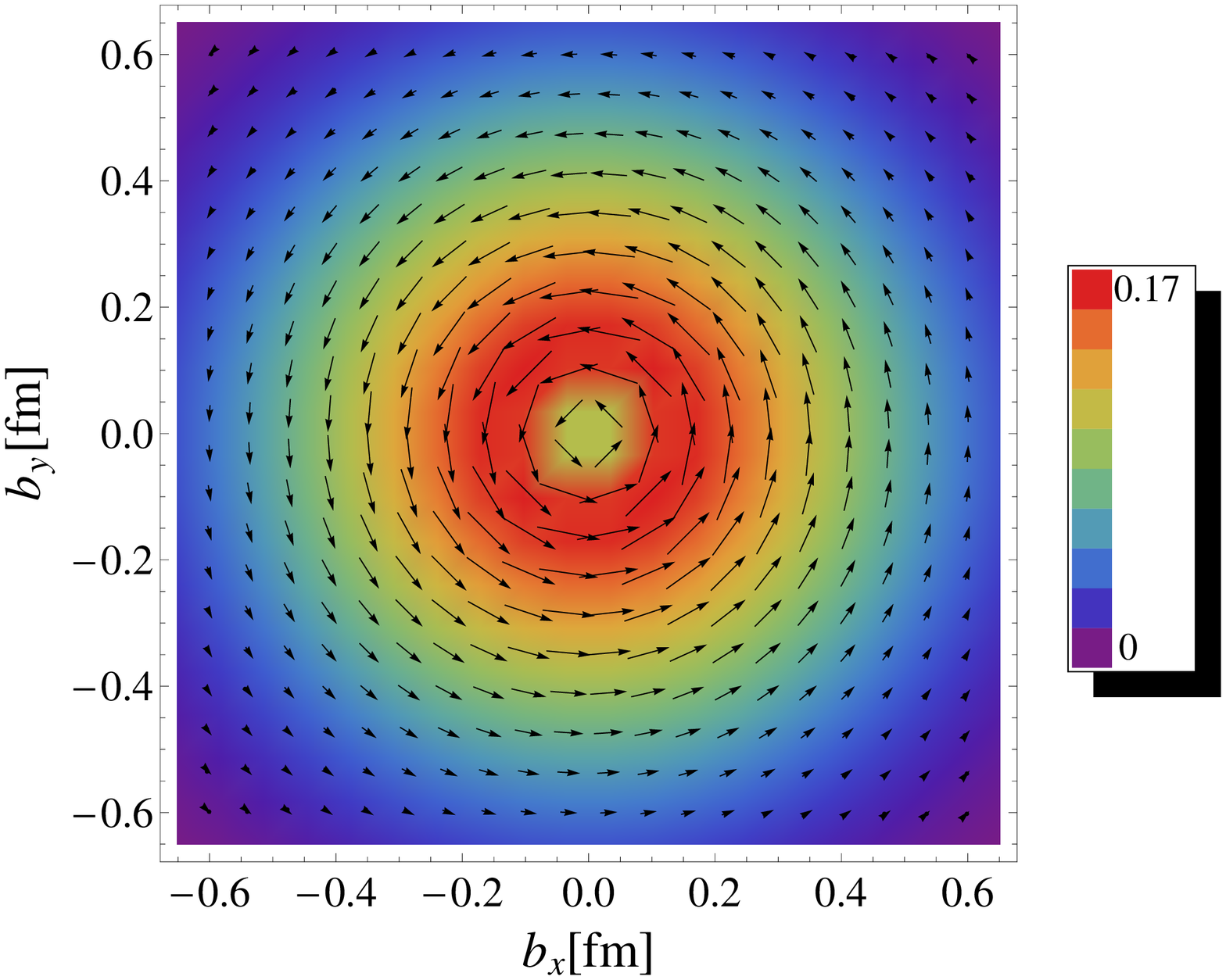}\hspace{1cm}
		\includegraphics[width=.42\textwidth]{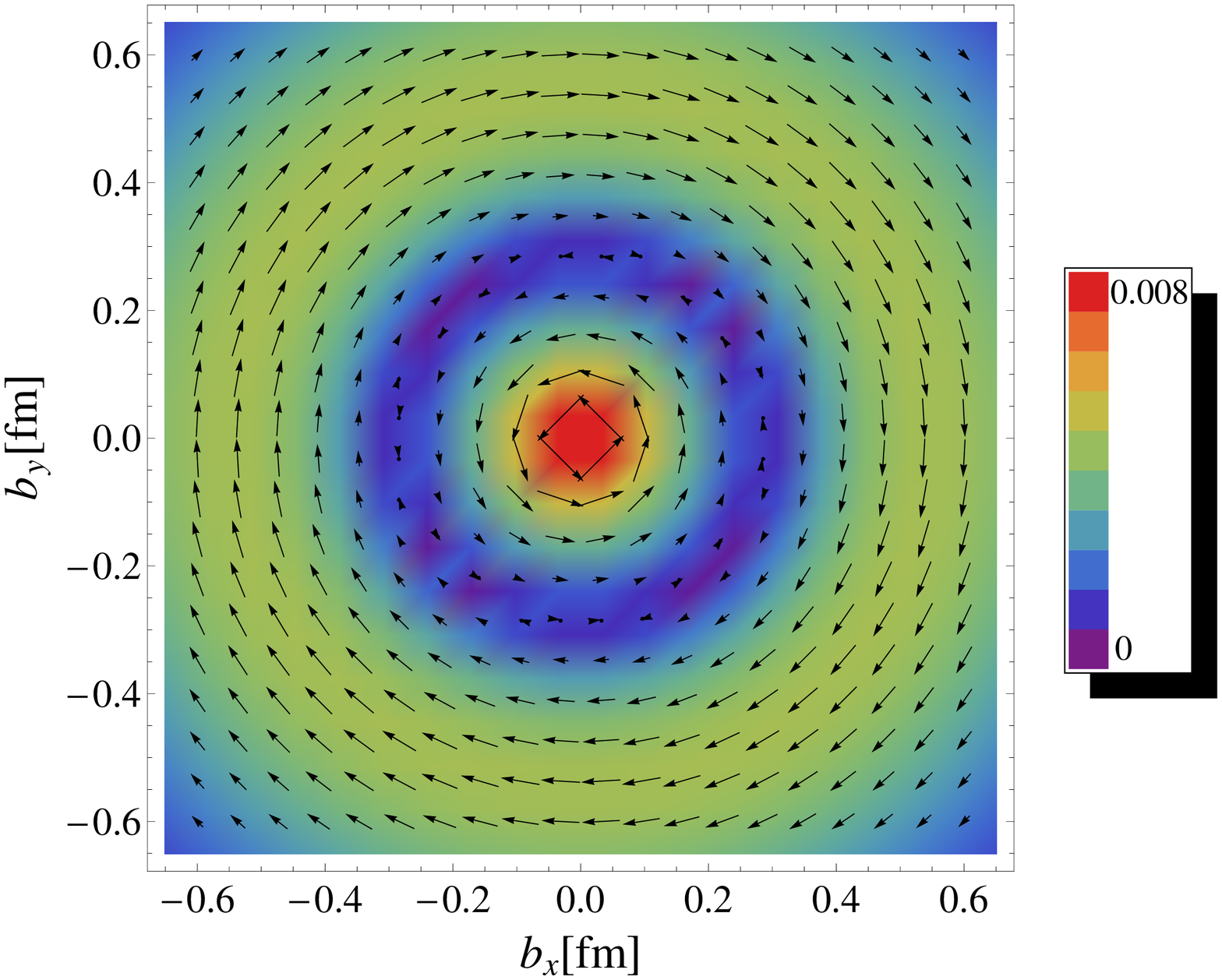}
	\caption{\footnotesize{The $x$-integrated distributions in impact-parameter space for unpolarized quarks in a longitudinally polarized proton (the proton spin points out of the plane). The upper panels show the distortion of the Wigner distribution, for a given transverse momentum $\vec k_\perp=k_\perp\,\vec e_y$ with $k_\perp=0.3$ GeV, induced by the proton polarization, and the lower panels show the distribution of the average quark transverse momentum. The left panels are for $u$ quarks and the right panels for $d$ quarks. These distributions have been obtained from the LFCQM~\cite{Lorce:2011kd,Lorce:2011ni}.}}\label{fig3}
	\vspace{-0.2 truecm}
\end{figure}

In Fig.~\ref{fig3}, the upper panels show the distortions in impact-parameter space for $u$ (left panel) and $d$ (right panel) quarks with fixed transverse momentum $\vec k_\perp=k_\perp\,\vec e_y$ and $k_\perp=0.3$ GeV. We observe a clear dipole structure in these distributions which indicates that the (quasi-)probability for finding the quark orbiting clockwise is not the same as for the quark orbiting anticlockwise, leading in average to a nonvanishing OAM.  
The lower panels of Fig.~\ref{fig3} describes the distribution in impact-parameter space 
of the  quark average transverse momentum in a longitudinally polarized nucleon
\begin{equation}
 \langle\vec k_\perp\rangle^q(\vec b_\perp)=\int\ud x\,\ud^2k_\perp\,\vec k_\perp\,\rho^{[\gamma^+]q}_{\Lambda\Lambda}(x,\vec k_\perp,\vec b_\perp,n).
\end{equation}
We observe that the average transverse momentum 
is always orthogonal to the impact-parameter vector
$\vec b_\perp$. This is not surprising since a nonvanishing radial component of the average transverse
momentum would indicate that the proton size and/or shape are changing.  
We can also clearly notice  that $u$ quarks tend to orbit anticlockwise inside the nucleon, corresponding to 
positive OAM aligned with the nucleon spin which is  pointing out of the figure. For the $d$
quarks, we see two regions. In the central region of the nucleon, $|\vec b_\perp|< 0.3$  fm, the $d$ quarks
tend to orbit anticlockwise like the $u$ quarks, while in the peripheral region, $|\vec b_\perp|>0.3$  fm, the
$d$ quarks tend to orbit clockwise, with a flip of the local net quark OAM. Note that such information about the OAM can not be accessed through GPDs and TMDs since none of them describe at leading twist the distortion in the distribution of unpolarized quarks due to the longitudinal polarization of the nucleon. 
 This is because one needs the correlation between $\vec b_\perp$ and $\vec k_\perp$ which is lost by integrating over $\vec b_\perp$ or $\vec k_\perp$.

The Wigner distributions were originally constructed as the quantum mechanical analogue
of the classical density operator in the phase space.   In particular, any matrix element of a quark operator can be rewritten as a phase-space integral of the corresponding classical quantity weighted by the Wigner distribution. It is therefore natural to define the quark OAM as follows~\cite{Lorce:2011kd}
\begin{equation}\label{OAMWigner}
l_z^q=\int\ud x\,\ud^2k_\perp\,\ud^2b_\perp\left(\vec b_\perp\times\vec k_\perp\right)_z\,\rho^{[\gamma^+]q}(\vec b_\perp,\vec k_\perp,x,+\vec e_z).
\end{equation}
Since the Wigner distribution involves in its definition a gauge link, it inherits a path dependence~\cite{Lorce:2012ce,Lorce:2012rr} . The simplest choice is a straight gauge link. In this case, Eq.~\eqref{OAMWigner} gives the kinetic OAM $L^q_z=l^{q,\text{straight}}_z$ associated with the quark OAM operator appearing in the Ji decomposition~\cite{Ji:1996ek,Ji:2012sj}  $\int\ud^3 r\,\overline \psi^q\gamma^+\,\mathbf r_\perp\times(-i\mathbf 
D_\perp)\psi^q$ , where $D_\mu=\partial_\mu-igA_\mu$ is the usual covariant derivative. According to the Ji's relation~\cite{Ji:1996ek}, this kinetic quark OAM can be extracted from the GPDs
\begin{equation}\label{ji-sumrule}
L^q_z=J^q_z-\frac{1}{2}\Delta\Sigma^q,
\end{equation}
with
\begin{eqnarray}
J^q_z&=&\frac{1}{2}\int^1_{-1}\ud x\left\{x\left[H^q(x,0,0)+E^q(x,0,0)\right]\right\},
\label{jisumrule}
\\
\Delta\Sigma^q&=&\int^1_{-1}\ud x \, \tilde H^q(x,0,0).
\end{eqnarray}
In order to connect the Wigner distributions to the TMDs, it is more natural to consider instead a staple-like gauge link consisting of two longitudinal straight lines connected at $x^-=\pm\infty$ by a transverse straight line. In this case, Eq.~\eqref{OAMWigner} gives the canonical OAM $\ell_z=l^{q,\text{staple}}_z$ associated with the quark OAM operator that appears in the Jaffe-Manohar decomposition in the $A^+=0$ 
gauge~\cite{Jaffe:1989jz,Lorce:2011ni,Hatta:2011ku}, i.e. $\int\ud^3 r\,\overline \psi^q\gamma^+\,\mathbf 
r_\perp\times(-i\boldsymbol\nabla_\perp)\psi^q$.

 \section{GPDs in impact-parameter space}
 \label{sec:3}
 
 In this section, we discuss a few examples of spin densities parametrized in terms of GPDs in impact-parameter space.
As outlined in the introduction, they can be obtained from the Wigner distributions after integration over the quark transverse momentum 
and can be   interpreted as probability densities of quarks with longitudinal momentum fraction $x$ and transverse location $\vec b_\perp$ with respect to the nucleon center of momentum~\cite{Burkardt:2005td}.
In Fig.~\ref{fig3} we show the results within the LFCQM~\cite{Boffi:2007yc,Pasquini:2007xz} in the case of unpolarized quarks in a transversely polarized nucleon.
This spin density is given by the sum of  a nucleon spin-independent contribution related to the GPD $H$ and a nucleon spin-dependent contribution from the GPD $E$, corresponding to monopole and dipole distributions in impact-parameter space, respectively.
The dipole 
contribution introduces a large distortion perpendicular to both the nucleon spin and the momentum of the proton, with opposite sign for $u$ and $d$ quarks.
Such a distortion reflects the large value of  the  anomalous 
magnetic moments $\kappa^{u,d}$. 
\begin{figure}[t]
\centerline{\psfig{file=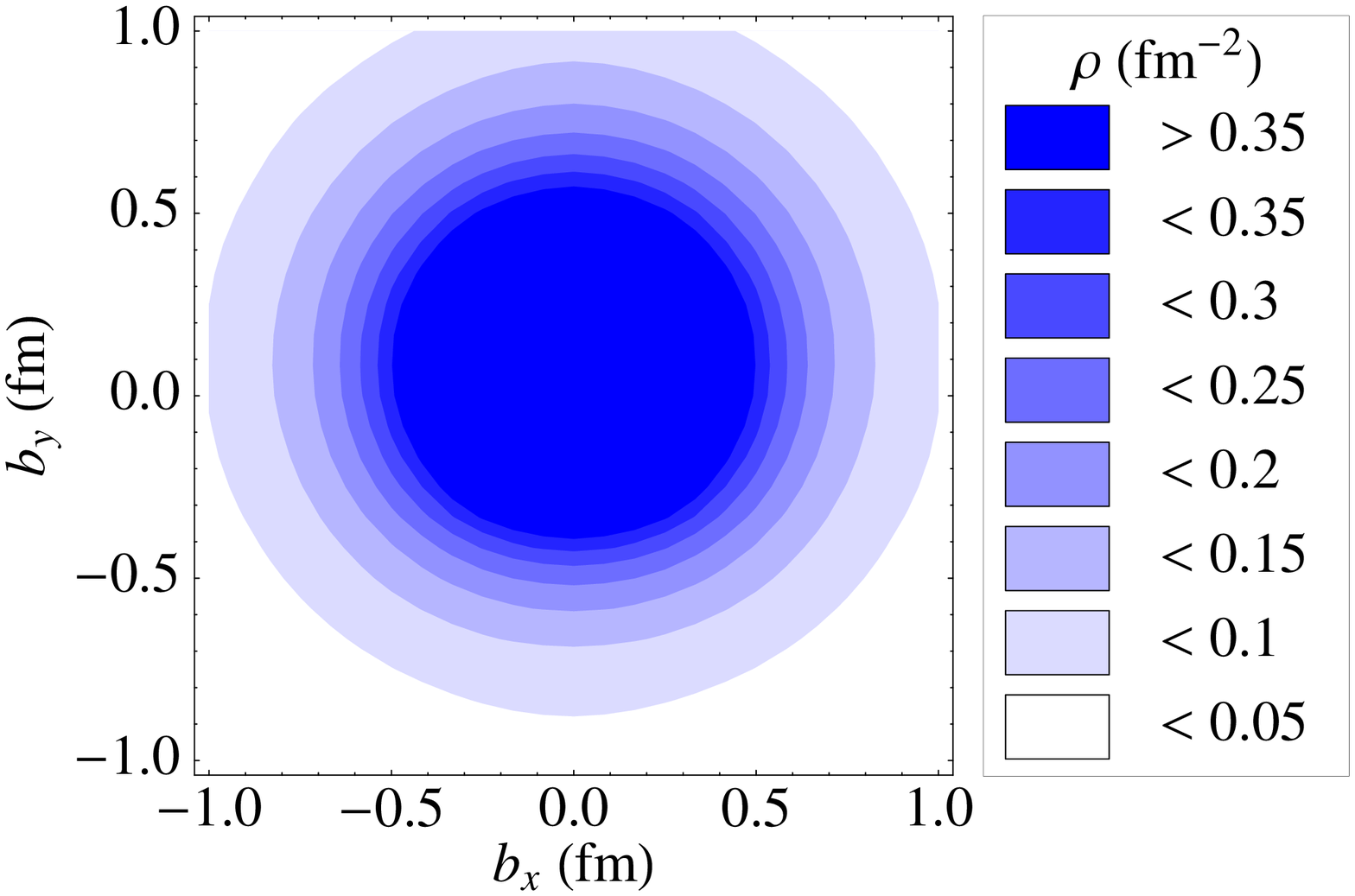,width=6.5 cm}{\psfig{file=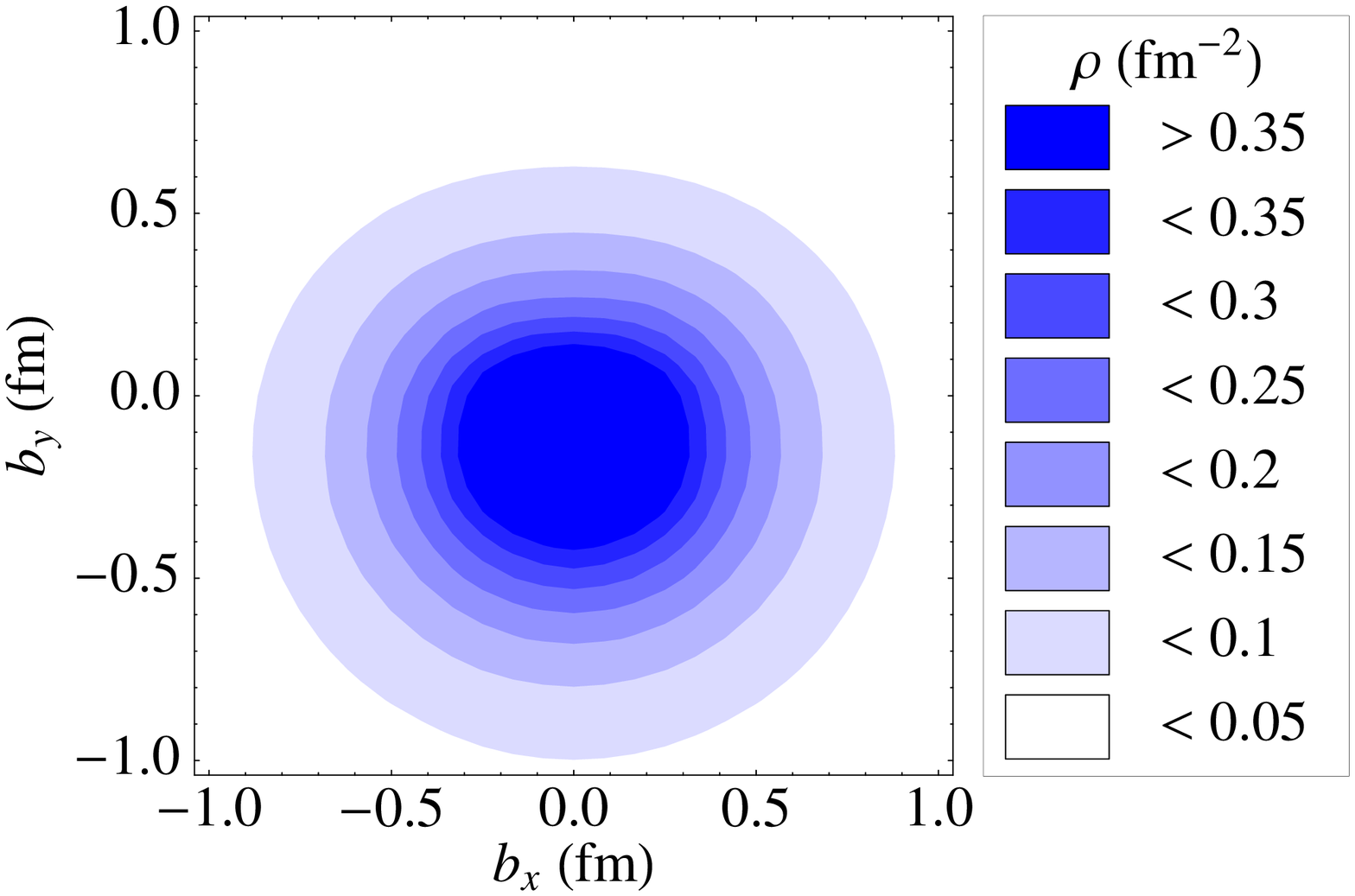,width=6.5 cm}}}
\vspace{-0.7 truecm}
\caption{The spin-densities
for unpolarized quarks in an (transversely) $\hat x$-polarized proton
for $u$ (left panel) and $d$ (right panel) quarks.
\protect\label{fig3}}
\end{figure}
With the present model, $\kappa^u=1.86$ and $\kappa^d=-1.57$, 
to be compared with the values $\kappa^u=1.673$ and $\kappa^d=-2.033$ 
derived from data. This effect can serve as a dynamical explanation of
 a non-vanishing Sivers function $f_{1T}^\perp$ 
which measures the correlation between the intrinsic quark transverse momentum 
and the transverse nucleon spin~\cite{Burkardt:2002ks}. 
This connection between the GPD $E$ and the Sivers function  has  recently been exploited in Ref.~\cite{Bacchetta:2011gx} to determine the total quark angular momentum from the Ji's relation~\eqref{jisumrule}, reconstructing the GPD $E$ in the collinear limit from  the available experimental 
information on $f_{1T}^\perp$  in SIDIS~\cite{Hermes05a,Alekseev:2008aa}.
Though this estimate is based on a  model-dependent relation, the consistency with constraints on the angular momentum arising from DVCS measurements~\cite{Airapetian:2008aa,Mazouz:2007aa}   is encouraging.

\section{TMDs in momentum space}
 \label{sec:4}

\begin{figure}[ht]
	\centering
		\includegraphics[width=6.cm]{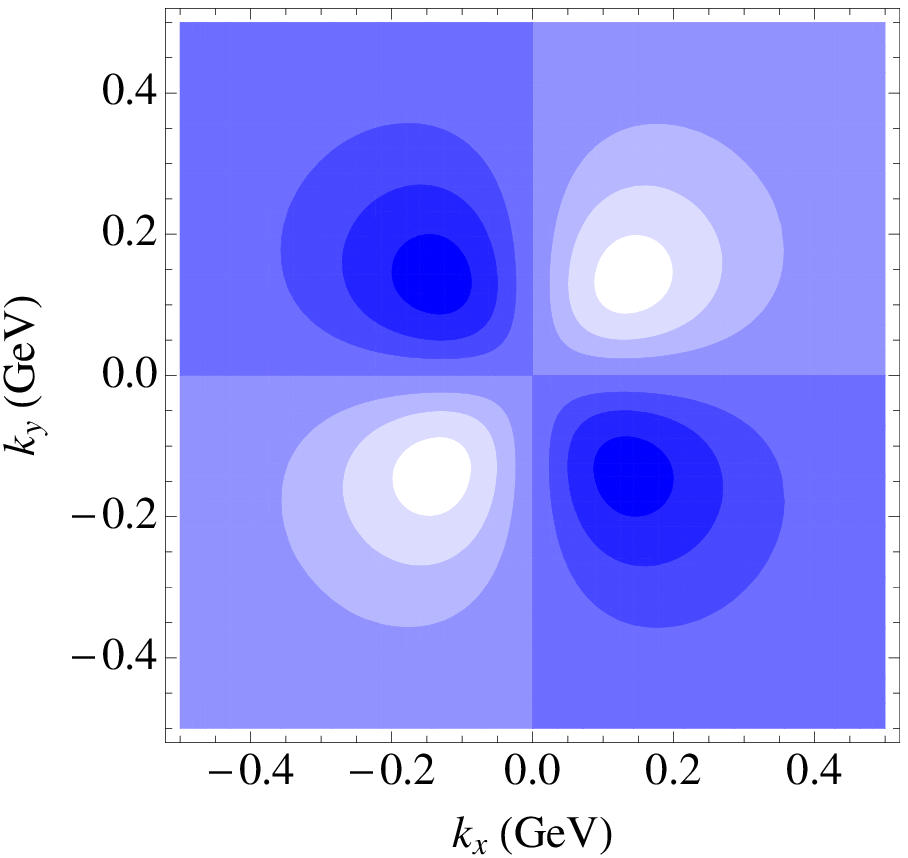}
\includegraphics[width=6.cm]{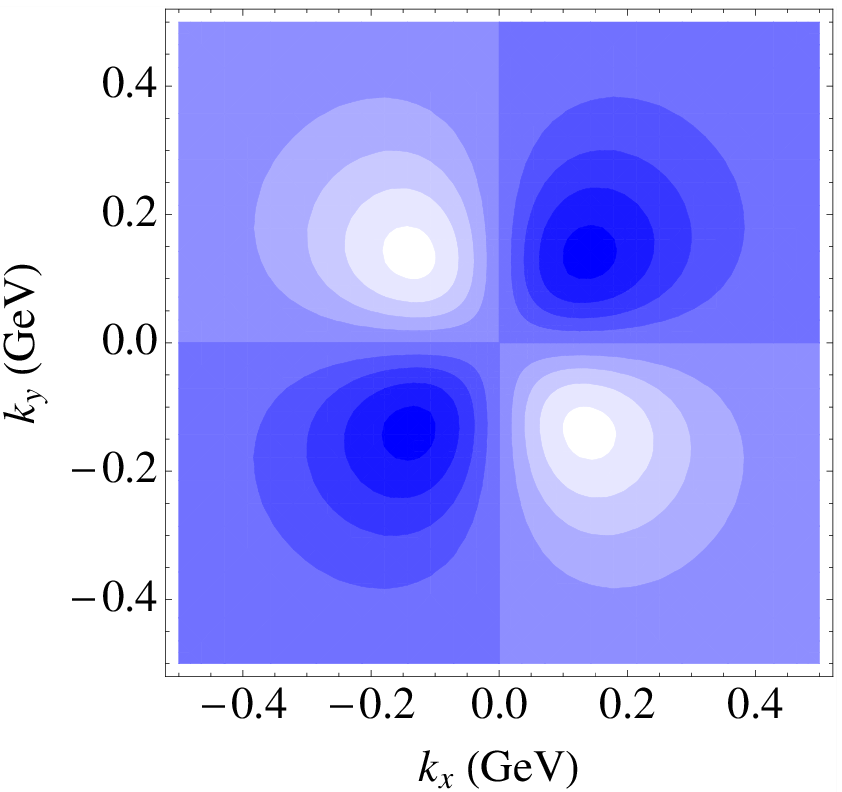}
\caption{Density of quarks in the $\uk_\perp$ plane
for net transverse polarization of quarks and proton in perpendicular 
directions.
The left and right  panel shows the results for up and down quarks, respectively.
\protect\label{fig4}}
\end{figure}

The eight leading-twist TMDs are a
natural extension of standard parton distribution from one to three dimensions in
momentum space, being function of both the longitudinal quark momentum fraction $x$ and the transverse momentum $\vec k_\perp$.
The knowledge of TMDs allow us to build tomographic images of the inner structure of the nucleon
in momentum space, complementary to the impact-parameter space tomography that can be achieved by studying
GPDs.
\\
The LFWF overlap representation of the TMDs has been explicitly derived in Refs.~\cite{Pasquini:2008ax,Pasquini:2010af,Brodsky:2010vs} and can be also obtained using the results of Sect.~\ref{sec:1} for the GTMDs in the forward limit $\Delta=0$. 
This representation is
well suited to illustrate the relevance of the different orbital angular momentum components
of the nucleon wave function, and provide an intuitive picture for the physical meaning of
the quark TMDs. Moreover, they can be regarded as initial input for phenomenological
studies for the semi-inclusive processes where quark TMDs play a very important role~\cite{Boffi:2009sh,Pasquini:2011tk}.
Most of the TMDs would simply vanish
in absence of quark OAM.
Recently, it has been suggested, on the basis of  some
quark-model calculations~\cite{She:2009jq,Avakian:2010br}, that the TMD
$h_{1T}^\perp$ may be related to the quark OAM:
\begin{equation}
\label{OAMpretzel}
\mathcal L_z^q=-\int\ud x\,\ud^2k_\perp\,\frac{k_\perp^2}{2M^2}\,h_{1T}^{\perp q}(x,k^2_\perp).
\end{equation}
However, Eq.~\eqref{OAMpretzel}  is not a  rigorous expression and holds only in a  specific  class of quark models. For a detailed discussion on the
the physical origin of this relation and the underlying
model assumptions for its validity we refer to~\cite{Lorce:2011kn}.
The  $h_{1T}^\perp$  TMD describes the distortion due to the transverse polarizations in perpendicular directions of the quark and the nucleon~\cite{Miller:2003sa}. In this case, the nucleon helicity flips in the direction opposite to the quark helicity, with a mismatch of two units for the orbital angular momentum of the initial and final LFWFs. The corresponding quadrupole structure  in the momentum space for both $u$ and $d$ quarks is shown in  Fig. \ref{fig4}, as obtained from the model of Ref.~\cite{Pasquini:2008ax}.

Finally, 
  in Table \ref{OAMtable}, we summarize the results from the LFCQM and the LF$\chi$QSM 
   for the quark OAM obtained from the Ji relation (Eq.~\eqref{ji-sumrule}), the Wigner distributions (Eq.~\eqref{OAMWigner}) and the $h_{1T}^\perp$  TMD (Eq.~\eqref{OAMpretzel}).

\begin{table}[th!]
\begin{center}
\caption{\footnotesize{Comparison between the Ji ($L^q_z$), Jaffe-Manohar ($\ell^{q,{\rm staple}}_z$) and TMD ($\mathcal L^q_z$) OAM in the LFCQM and the LF$\chi$QSM for $u$-, $d$- and total ($u+d$) quark contributions.}}\label{OAMtable}
\begin{tabular}{@{\quad}c@{\quad}|@{\quad}c@{\quad}c@{\quad}c@{\quad}|@{\quad}c@{\quad}c@{\quad}c@{\quad}}\hline
Model&\multicolumn{3}{c@{\quad}|@{\quad}}{LFCQM}&\multicolumn{3}{c@{\quad}}{LF$\chi$QSM}\\
$q$&$u$&$d$&Total&$u$&$d$&Total\\
\hline
$L^q_z$&$0.071$&$~~0.055$&$0.126$&$-0.008$&$~~0.077$&$0.069$\\
$\ell^q_z$&$0.131$&$-0.005$&$0.126$&$~~0.073$&$-0.004$&$0.069$\\
$\mathcal L^q_z$&$0.169$&$-0.042$&$0.126$&$~~0.093$&$-0.023$&$0.069$\\
\hline
\end{tabular}
\end{center}
\end{table}
As expected in a pure quark model, all the definitions give the same value for the total quark OAM, with nearly twice more net quark OAM in the LFCQM than in the LF$\chi$QSM. The difference between the various definitions appears in the separate quark-flavour contributions. Note in particular that unlike the LFCQM, the LF$\chi$QSM predicts a negative sign for the $u$-quark OAM in agreement with lattice calculations~\cite{Hagler:2007xi}. 
It is surprising that $\ell^q_z\neq L^q_z$ since it is generally believed that the Jaffe-Manohar and Ji's OAM should coincide in absence of gauge degrees of freedom. Note that a similar observation has also been made in the instant-form version of the $\chi$QSM~\cite{Wakamatsu:2005vk}.
On the other hand, the individual
quark contributions to the OAM
obtained from Eq.~\eqref{OAMpretzel} do not correspond
to the intrinsic quark orbital angular momentum, and therefore do not coincide with the results for  $\ell^q_z$.
The two calculations agree only for the total
OAM, since in the sum over the individual quark contributions
the spurious terms due to the transverse centre
of momentum cancel out.

\section{Conclusions}
In this work we presented a study of GTMDs, which parametrize the fully-unintegrated  quark-quark correlators with the quark fields are taken at the same light-front time. By taking specific limits or projections
of these GTMDs, they yield PDFs, TMDs, GPDs, FFs, and charges, accessible in
various inclusive, semi-inclusive, exclusive, and elastic scattering processes. The GTMDs
therefore provide a unified framework to simultaneously model these different observables.
We discussed a first step in this modeling, by considering a light-front wave function (LFWF) 
overlap representation of the GTMDs and by restricting ourselves to the 3Q Fock components
in the nucleon LFWF. At twist-two level, we studied the most general transition
which the active quark light-front helicity can undergo in a polarized nucleon, corresponding
to the general helicity amplitudes of the quark-nucleon system. We develop a formalism
which is quite general and can be applied to many quark models as long as the nucleon
state can be represented in terms of 3Q without mutual interactions. 
By Fourier transform in the transverse space of the GTMDs we obtain the Wigner distributions which provide the multidimensional images of the quark distributions in the phase space. 
In particular,  we discussed results for the Wigner distributions of
unpolarized quarks in a longitudinally polarized nucleon that allow us to calculate the phase-space average of the quark OAM. 
Other ways to access information about the quark OAM from GPDs and TMDs have been also discussed, comparing the corresponding results obtained 
within a light-front constituent quark model and the light-front chiral quark-soliton model.
\section*{Acknowledgments}

 This work was supported in part by the Research Infrastructure Integrating Activity ÒStudy of Strongly Interacting MatterÓ (acronym HadronPhysic3, Grant Agreement n. 283286) under the Seventh Framework Programme of the European Community, by the Italian MIUR through the PRIN 2008EKLACK ÒStructure of the nucleon: transverse momentum, transverse spin and orbital angular momentumÓ, and by the P2I (ÒPhysique des deux InfinisÓ) network.


\begin{thebibliography}{}

\bibitem{Meissner:2009ww} 
  S.~Meissner, A.~Metz and M.~Schlegel,
  JHEP {\bf 0908}, (2009) 056.

  \bibitem{Meissner:2008ay}
  S.~Meissner, A.~Metz, M.~Schlegel and K.~Goeke,
  JHEP {\bf 0808}, (2008) 038.
\bibitem{Lorce:2011dv} 
  C.~Lorc\'e, B.~Pasquini and M.~Vanderhaeghen,
  JHEP {\bf 1105}, (2011) 041.

\bibitem{Ji:2003ak}
  X.~d.~Ji,
  Phys.\ Rev.\ Lett. {\bf 91}, (2003) 062001.

\bibitem{Belitsky:2003nz}
  A.~V.~Belitsky, X.~d.~Ji and F.~Yuan,
   Phys.\ Rev.\ D {\bf 69}, (2004) 074014.

\bibitem{Belitsky:2005qn}
  A.~V.~Belitsky and A.~V.~Radyushkin,
  Phys.\ Rept.  {\bf 418}, (2005) 1.

\bibitem{Lorce:2011kd} 
  C.~Lorc\'e and B.~Pasquini,
  Phys.\ Rev.\  D {\bf 84}, (2011) 014015.

\bibitem{Lorce:2011ni} 
  C.~Lorc\'e, B.~Pasquini, X.~Xiong and F.~Yuan,
  Phys.\ Rev. \  D {\bf 85}, (2012) 114006.
  
  
\bibitem{Diehl:2000xz}
  M.~Diehl, T.~Feldmann, R.~Jakob and P.~Kroll,
  Nucl.\ Phys. B {\bf 596}, (2001) 33
  [Erratum-ibid.\ B  {\bf 605}, (2001) 647].

\bibitem{Brodsky:2000xy}
  S.~J.~Brodsky, M.~Diehl and D.~S.~Hwang,
  Nucl.\ Phys.  B {\bf 596}, (2001) 99.

  \bibitem{Boffi:2002yy}
  S.~Boffi, B.~Pasquini and M.~Traini,
  Nucl.\ Phys.\ B  {\bf 649}, (2003) 243.

\bibitem{Boffi:2003yj}
  S.~Boffi, B.~Pasquini and M.~Traini,
  Nucl.\ Phys.\   B {\bf 680}, (2004) 147.

\bibitem{Pasquini:2005dk}
  B.~Pasquini, M.~Pincetti and S.~Boffi,
  Phys.\ Rev.\  D {\bf 72}, (2005) 094029;
  Phys.\ Rev.\ D {\bf 76}, (2007) 034020.

  \bibitem{Petrov:2002jr}
  V.~Y.~Petrov and M.~V.~Polyakov,
  arXiv:hep-ph/0307077.

\bibitem{Diakonov:2005ib}
  D.~Diakonov and V.~Petrov,
  Phys.\ Rev.\  D {\bf 72}, (2005) 074009.

\bibitem{Lorce:2007as}
 C.~Lorc\'e,
  Phys.\ Rev.\   D {\bf 78}, (2008) 034001.


\bibitem{Lorce:2007fa}
  C.~Lorc\'e,
  Phys.\ Rev.\   D {\bf 79}, (2009) 074027.
  \bibitem{Lorce:2011zta}
  C.~Lorc\'e and B.~Pasquini,
  Phys.\ Rev.\ D {\bf 84}, (2011) 034039.
  
\bibitem{Pasquini:2012jm}
  B.~Pasquini and C.~Lorc\'e,
  arXiv:1203.5006 [hep-ph].
  
  
  
  \bibitem{Pasquini:2007iz}
  B.~Pasquini and S.~Boffi,
  Phys.\ Rev.\  D {\bf 76}, (2007) 074011.
  %
      \bibitem{Lorce:2012ce} 
     C.~Lorc\'e, 
  arXiv:1210.2581 [hep-ph].
  
 
\bibitem{Lorce:2012rr} 
  C.~Lorc\'e, 
  arXiv:1205.6483 [hep-ph].
  
\bibitem{Ji:1996ek}
  X.~D.~Ji, 
  Phys.\ Rev.\ Lett.\  {\bf 78}, (1997)  610. 
\bibitem{Ji:2012sj} 
  X.~Ji, X.~Xiong and F.~Yuan,
  Phys.\ Rev.\ Lett.\  {\bf 109}, (2012) 152005.

\bibitem{Jaffe:1989jz} 
  R.~L.~Jaffe and A.~Manohar,
  Nucl.\ Phys.\ B {\bf 337}, (1990) 509.

\bibitem{Hatta:2011ku} 
  Y.~Hatta, 
   Phys.\ Lett.\ B {\bf 708}, (2012) 186    

\bibitem{Burkardt:2005td}
  M.~Burkardt,
  Int.\ J.\ Mod.\ Phys.\  A {\bf 21}, (2006) 926;
  Int.\ J.\ Mod.\ Phys.\  A {\bf 18}, (2003) 173;
  Phys.\ Rev.\  D {\bf 62}, (2000) 071503
  [Erratum-{\em ibid.}\   {\bf 66}, (2002) 119903].


\bibitem{Boffi:2007yc}
  S.~Boffi and B.~Pasquini,
  Riv.\ Nuovo Cim. {\bf 30}, (2007) 387.
  \bibitem{Pasquini:2007xz}
  B.~Pasquini and S.~Boffi,
  Phys.\ Lett.\ B {\bf 653}, (2007) 23.
  \bibitem{Burkardt:2002ks}
  M.~Burkardt,
  Phys.\ Rev.\ D {\bf 66}, (2002) 114005;
  M.~Burkardt and D.~S.~Hwang,
  Phys.\ Rev.\ D {\bf 69}, (2004) 074032.
  
\bibitem{Bacchetta:2011gx}
  A.~Bacchetta and M.~Radici,
  Phys.\ Rev.\ Lett.\  {\bf 107}, (2011) 212001.
  
\bibitem{Hermes05a}
A.~Airapetian {\it et al.\/} (Hermes Collaboration), Phys. Rev. Lett. {\bf 94}, (2005) 012002.

\bibitem{Alekseev:2008aa}
  M.~Alekseev {\it et al.}  [COMPASS Collaboration],
  Phys.\ Lett.\ B {\bf 673}, (2009) 127.

  \bibitem{Airapetian:2008aa}
  A.~Airapetian {\it et al.}  [HERMES Collaboration],
  JHEP {\bf 0806}, (2008) 066.
  
  
    \bibitem{Mazouz:2007aa}
  M.~Mazouz {\it et al.}  [Jefferson Lab Hall A Collaboration],
  Phys.\ Rev.\ Lett.\  {\bf 99}, (2007) 242501.
            
              \bibitem{Pasquini:2008ax}
  B.~Pasquini, S.~Cazzaniga and S.~Boffi,
  Phys.\ Rev.  D  {\bf 78}, (2008) 034025.

\bibitem{Pasquini:2010af}
  B.~Pasquini and F.~Yuan,
  Phys.\ Rev.  D {\bf 81}, (2010) 114013.
    
\bibitem{Brodsky:2010vs}
  S.~J.~Brodsky, B.~Pasquini, B.~-W.~Xiao and F.~Yuan,
  Phys.\ Lett.\ B {\bf 687} (2010) 327.

  
  \bibitem{Boffi:2009sh}
  S.~Boffi, A.~V.~Efremov, B.~Pasquini and P.~Schweitzer,
  Phys.\ Rev.\  D {\bf 79}, (2009) 094012.
\bibitem{Pasquini:2011tk}
  B.~Pasquini and P.~Schweitzer,
  Phys.\ Rev.\ D {\bf 83}, (2011) 114044.


\bibitem{She:2009jq}
  J.~She, J.~Zhu and B.~-Q.~Ma,
  Phys.\ Rev.\ D {\bf 79}, (2009) 054008.
  
  \bibitem{Avakian:2010br}
  H.~Avakian, A.~V.~Efremov, P.~Schweitzer and F.~Yuan,
  Phys.\ Rev.\ D {\bf 81}, (2010) 074035.
  
  
  \bibitem{Lorce:2011kn} 
  C.~Lorc\'e and B.~Pasquini, 
  Phys.\ Lett.\ B {\bf 710}, (2012) 486 
  
  
\bibitem{Miller:2003sa}
  G.~A.~Miller,
  Phys.\ Rev.\ C {\bf 68}, (2003) 022201.
  
\bibitem{Hagler:2007xi}
  Ph.~H\"agler {\it et al.}  [LHPC Collaborations], 
  Phys.\ Rev.\  D {\bf 77}, (2008) 094502.

\bibitem{Wakamatsu:2005vk}
  M.~Wakamatsu, H.~Tsujimoto, 
  Phys.\ Rev.\  D {\bf 71}, (2005) 074001;
  M.~Wakamatsu, 
  Eur.\ Phys.\ J.\  A {\bf 44}, (2010) 297.
          
  
\end{thebibliography}
\end{document}